\documentclass[doublecol,figures]{epl2} 

\usepackage{epsf}
\usepackage{color}
\usepackage{amssymb}
\usepackage{amsfonts}
\usepackage{amsmath}

\newcommand{\half}{\mbox{\small $\frac{1}{2}$}}
\newcommand{\bra}[1]{\left\langle #1\right|}
\newcommand{\ket}[1]{\left|#1\right\rangle }

\title{Engineering fidelity echoes in Bose-Hubbard Hamiltonians}
\author{
Joshua D. Bodyfelt\inst{1}, Moritz Hiller\inst{1,2,3} \and Tsampikos Kottos\inst{1,2}
}
\institute{
    \inst{1} Wesleyan University, Department of Physics - 265 Church St., Middletown, CT 06459, USA \\
    \inst{2} MPI for Dynamics and Self-Organization - Bunsenstra\ss e 10, D-37073 G\"ottingen, Germany\\
    \inst{3} Fakult\"at f\"ur Physik, Universit\"at G\"ottingen - Friedrich-Hund-Platz 1, D-37077 G\"ottingen, Germany
}


\abstract{
We analyze the fidelity decay for a system of interacting bosons described by a Bose-Hubbard 
Hamiltonian. We find echoes associated with "non-universal" structures that dominate the 
energy landscape of the perturbation operator. Despite their classical origin, these echoes 
persist deep into the quantum (perturbative) regime and can be described by an improved random 
matrix modeling. In the opposite limit of strong perturbations (and high enough energies), 
classical considerations reveal the importance of self-trapping phenomena in the echo 
efficiency.}
\begin{document}
\maketitle

%%%%%%%%%%%%%%%%%%%%%%%%%%%%     Introduction     %%%%%%%%%%%%%%%%%%%%%%%%%%%%%%%%%%%%%
While the physics of the last century was mainly characterized by great advances in 
understanding the properties of one-particle systems, recent experimental developments 
have put the effects of interacting bosons at the top of the research agenda. Among the 
experimental realizations are systems as diverse as micro-mechanical arrays~\cite{SHSICG03}, 
coupled Josephson Junctions~\cite{TMO00}, and Bose-Einstein Condensates (BEC) loaded in 
optical lattices~\cite{JBCGZ98}, the latter being suggested as viable candidates for 
quantum computation.

Apart from their technological applications, these systems offer unique opportunities to 
experimentally investigate fundamental problems of quantum mechanics. One of the main questions 
is the manipulation of coherence and the stability of complex quantum dynamics under external 
(environmental) perturbations. A well-established measure for the latter is the so-called {\sl 
fidelity} (also known as the Loschmidt Echo). It was first introduced by Peres \cite{P84} who 
used fidelity to study quantum-classical correspondence and identify traces of classical (chaotic 
or integrable) dynamics in quantized systems. Recently, fidelity has been used to quantify the 
corruption of quantum information~\cite{QCompBook}, in the framework of quantum computation. 

In this Letter, we study the fidelity decay of (ultra-) cold atoms loaded in an optical lattice 
(OL) or coupled micro-traps subject to perturbations of the coupling: $k_0\rightarrow k_0+\delta 
k$. In the context of OLs this perturbation is achieved by adjusting the intensity of the laser 
beams that create the lattice. Such systems are described by the Bose-Hubbard Hamiltonian (BHH), 
which in second quantization reads
\begin{equation}
\hat{H}=\frac{U}{2}\sum_{i=1}^{f}{\hat n}_i ({\hat n}_i-1) -
k\sum_{\langle i,j\rangle}\hat{b}_{i}^{\dagger}\hat{b}_{j};\quad \hbar=1,
\label{QMBHH3}
\end{equation}
where $f$ is the number of sites and $\langle i,j\rangle$ denotes summation over adjacent sites and $U=4 \pi
\hbar^2a_sV_{\rm eff}/m$ describes the interaction between two atoms on a single site ($V_{
\rm eff}$ is the effective mode volume of each site, $m$ is the atomic mass, and $a_s$ is the 
$s$-wave atomic scattering length). The operators ${\hat n}_i=\hat{b}_i^{\dagger}\hat{b}_i$ 
count the number of bosons at site ~$i$; the annihilation and creation operators $\hat{b}_i$ 
and $\hat{b}_i^{\dagger}$ obey the canonical commutation relations $[\hat{b}_i,\hat{b}_j^{
\dagger}]=\delta_{i,j}$. In the context of Josephson Junction arrays, $k$ is given by the 
Josephson energy $E_J$ while $U$ accounts for the Coulomb interaction of the charged bosons
~\cite{BFS05}. 

We will focus our presentation on the trimer $f=3$. This is the minimum BHH model that contains 
all the generic ingredients~\cite{noteonNiu1} (like classical chaotic dynamics~\cite{SLE85}) 
of large BHH lattices and therefore often is used as a prototype model~\cite{FGS89}. Our main 
results are summarized in fig.~1 where we plot the fidelity $F(t)$ as function of time for 
various perturbation strengths $\delta k$. The new striking feature is the appearance of echoes 
at multiples of $t_{\rm echo}$. By analyzing the energy landscape of the perturbation operator, 
we are able to identify $t_{\rm echo}$ and {\it control the echo efficiency} by an appropriate 
choice of the initial preparation. For moderate perturbations $\delta k$, an improved RMT 
modeling that incorporates the semiclassical structures of the perturbation operator can 
reproduce these echoes, while for larger perturbations we rely on semiclassical considerations. 
We show that the trajectories leading to fidelity echoes becomes more abundant at high 
energies, in contrast to recent experimental results on echo spectroscopy on ultra-cold 
atoms in atom-optics billiards \cite{AGKD04} where it was found that the echoes do not 
survive in the strong perturbation limit. Our analysis indicates that this is due to self-trapping 
phenomena~\cite{TK88} and reflects a generic property of interacting bosons loaded on a
lattice.

%----------------------------------------------------------------------------------------------------
\begin{figure}[t]
\onefigure[width=\columnwidth,keepaspectratio, clip]{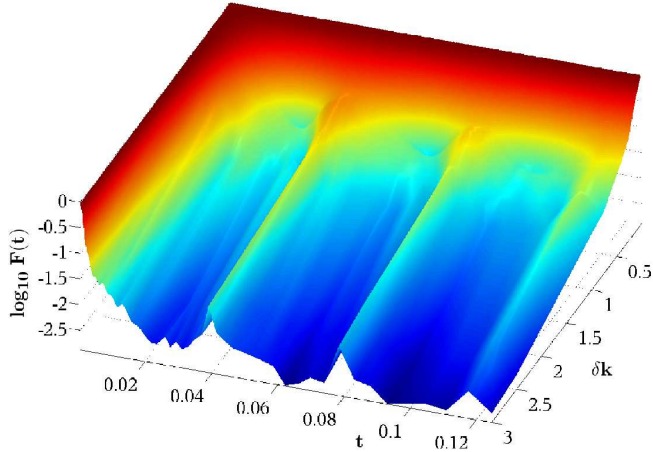}
\caption{(Color online) Parametric evolution of the fidelity $F(t)$ for different perturbation strengths 
$\delta k$ as a function of time. The fidelity exhibits echoes at multiples of $t=t_{\rm echo}$.
Here the Heisenberg time is $t_H\simeq 0.55$.}
\label{fig1}
\end{figure}
%----------------------------------------------------------------------------------------------------

%%%%%%%%%%%%%%%%%%%%     quantum framework and perturbation theory   %%%%%%%%%%%%
The fidelity is defined as the overlap of two states obtained after time-evolving the same 
initial preparation $\ket{\Psi_0}$ with two slightly perturbed Hamiltonians 
\begin{equation}
F(t) = \vert \bra{\Psi_0} e^{i \hat{H}_2 t} e^{-i \hat{H}_1 t} \ket{\Psi_0} \vert^2
\label{QMFid}
\end{equation}
In our analysis, we consider that $\hat{H}_{1,2}=\hat{H}_0\mp\delta k\hat{B}$. The unperturbed 
Hamiltonian $\hat{H}_0$ is given by eq.~(\ref{QMBHH3}) with $k=k_0$. In the simulations below
we will use $k_0\approx 15,\tilde{U}\approx 280$. The number of particles in the simulations is
$N=230$. The coupling operator $\hat{B}$ is defined as
\begin{equation}
{\hat B} = \sum_{\langle i,j\rangle}\hat{b}_{i}^{\dagger}\hat{b}_{j}.
\label{Boper}
\end{equation}
This perturbation is similar to a momentum boost. Such a perturbation has recently been 
investigated \cite{PBHJ06} in the context of the fidelity, where it was found that the fidelity 
freezes at some finite value, as long as the boost is not too large.

Quantum mechanically, we work in the ${\hat H}_0$ eigenbasis. The initial preparation $\ket{
\Psi_0}$ is chosen to be either a single eigenstate of ${\hat H}_0$ or a Gaussian superposition 
of them centered at energy $E_0$. A fixed assumption of this work is that the perturbation is {\sl 
classically} small $\delta k \ll \delta k_{\rm cl}$, \textit{i.e.}, the corresponding classical Hamiltonians 
${\cal H}_0$, ${\cal H}_1$ and ${\cal H}_2$ are generators of classical dynamics of the same 
nature. The latter is obtained from Eq.~(\ref{QMBHH3}) in the limit $N\gg 1$, and reads~\cite{HKG06}
\begin{equation}
 \tilde{\cal H} = {\cal H}/{N\tilde{U}}=\half\sum_{i=1}^{f}I_{i}^{2}-\lambda\sum_{\langle i,j\rangle}
{\sqrt {I_{i}I_{j}}} e^{i(\varphi_j-\varphi_i)}
\label{BHHCL}
\end{equation}
where $\{I_i,\phi_i\}$ are the canonical action-angle variables, and $\tilde{U} = NU$. The 
classical dynamics is governed by the dimensionless ratio $\lambda =k/\tilde{U}$~\cite{SLE85,
HKG06}. The other dimensionless parameter that affects the classical motion is the energy 
${\tilde E}={\tilde {\cal H}}$. 

An important observable is the generalized force ${\tilde{\cal F}}({\tilde t})\equiv-(\partial 
{\tilde {\cal H}}/\partial \lambda) = \sum_{\langle i,j\rangle} {\sqrt {I_{i}I_{j}}} \exp[i 
(\varphi_j-\varphi_i)] $ where ${\tilde t}={\tilde U} t$ is a rescaled time. We 
can characterize its fluctuations by the autocorrelation function $C({\tilde t})$. Its 
Fourier transform $\tilde{C} ({\tilde \omega})$ is the power spectrum of the fluctuations. 
Using a simple semiclassical recipe, we can connect $\tilde{C} ({\tilde \omega})$ with
the band-profile of the perturbation matrix ${\bf B}$~\cite{FP86}:
\begin{equation} 
\label{pspect}
\sigma^2\equiv \langle|{\bf B}_{nm}|^2\rangle_E 
\approx (N^2/{\tilde U})\Delta \cdot \tilde{C}(\omega_{nm})/2\pi
\end{equation}
where we use the basis which is determined by ${\hat H}_0$. Above $\langle\cdots\rangle_E$ 
indicates an averaging over an energy window $\delta E$ around some energy $E$, while 
$\Delta \sim{\tilde U}/N$~\cite{HKG06} is the mean level spacing. In cases of chaotic 
dynamics $\tilde{C}(\omega)$ has a cutoff frequency $\omega_{c} = 2\pi/\tau_{c}$ which is 
characterized by a finite correlation time $\tau_{c}$. Equation~(\ref{pspect}) implies that 
${\bf B}$ is a banded matrix with a bandwidth $\Delta_b=\hbar\omega_c$. 

Using Linear Response Theory~\cite{PZ02,GPSZ06} one can obtain the following standard expression
for the fidelity:
\begin{equation}
F(t) \simeq 1 - \delta k^2 C(t)+\cdots \approx \exp(-\delta k^2 C(t)),
\label{LRT}
\end{equation}
where $C(t)= \int_0^t\int_0^t C(t'-t'')dt'dt''$. This expression implies a short time Gaussian
decay $F(t)=\exp[-8 C(0)\times(\delta k \cdot t)^2]$, which evolves into a long time Gaussian decay 
$F(t)=\exp[-(\sigma\delta k)^2 t^2]$ for quantum mechanically small perturbations $\delta k\leq 
\delta k_{\rm qm} \sim \Delta/\sigma$~\cite{GPSZ06,JSB01,BC02,CLMPV02}. The decay of $F(t)$ for
small perturbations $\delta k \leq \delta k_{\rm qm}$, is the same irrespective of the nature 
(integrable or chaotic) of the underlying classical dynamics. 

\begin{figure}[t]
\onefigure[width=\columnwidth,keepaspectratio,clip]{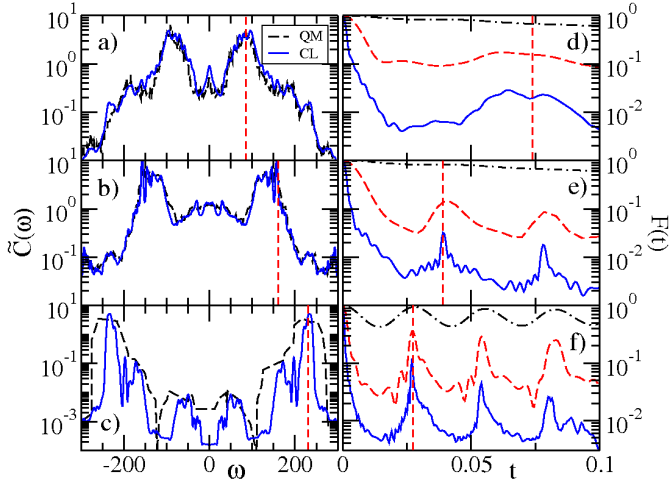}
\caption{(Color online) 
The quantum band-profile $\langle|{\bf B}_{nm}|^2\rangle{\tilde U}/ N^2 \Delta$ 
and classical power spectrum $\tilde{C}(\omega)$ (drawn with dashed and solid lines 
respectively) at various eigenvalues ${\tilde E}_n^{(0)}$ of the reference
Hamiltonian $H_0$: a) ${\tilde E}_n^{(0)}=0.22$, b) ${\tilde E}_n^{(0)}=0.26$, c) ${\tilde E}_n^{(0)}=0.39$.
In panels d)-f) we plot the corresponding fidelity $F(t)$ for three representative 
values $\delta k<\delta k_{\rm qm}$ 
(dash-dotted line, deep perturbative regime), $\delta k_{\rm qm}<\delta k<\delta k_{\rm prt}$ 
(dashed line, perturbative regime), and $\delta k>\delta k_{\rm prt}$ (solid line, non-perturbative regime). 
The respective numerical values are for d) and e) $\delta k = 0.1,0.5,2.5$, and for f) $\delta k = 0.5,2.5,7.5$. 
The initial preparation is an eigenstate of $H_0$. 
The vertical dashed lines in sub-figures d)-f) denote the revival time $t_{echo}$ 
while in a)-c) they denote the corresponding frequency $\omega^*=2\pi/t_{echo}$.}
\label{fig2}
\end{figure}

If the band-profile of the perturbation operator ${\bf B}$ were flat, we would expect~\cite{GPSZ06,
JSB01,BC02,CLMPV02} an exponential decay $F(t)\sim \exp[-2\Gamma t]$ for stronger perturbations 
$\delta k_{\rm qm}\leq \delta k \leq \delta k_{\rm prt} \sim \tilde{U}/N$. The rate $\Gamma \sim 
(\sigma \delta k)^2/\Delta$ is given by the width of the Local Density of States (LDoS)~\cite{HKG06}. 
In the perturbative regime (\textit{i.e.} $\delta k \leq \delta k_{\rm prt}$), traditional RMT considerations 
\cite{JSB01,GPSZ06} can describe the fidelity decay for chaotic systems \cite{GPSZ06,JSB01,CK01}. For 
even stronger perturbations $\delta k > \delta k _{\rm prt}$, we enter the semiclassical regime 
\cite{JSB01,BC02,CLMPV02}. This identification of the non-perturbative regime with the semiclassical 
limit can be easily understood if we realize that $\delta k_{\rm prt}\sim {\tilde U}/N\rightarrow 
0$ in the limit $N\gg1$ (while ${\tilde U}={\rm const.}$~ \cite{HKG06}). Then for any fixed 
perturbation $\delta k$, eventually $\delta k \gg \delta k_{\rm prt}$. In this regime (and provided
that the classical dynamics is chaotic) the fidelity decay is exponential $F(t)\sim \exp(-\gamma 
t)$, with a rate given by the classical Lyapunov exponent $\gamma$ \cite{note3}. Our numerical 
data reported in figs.~\ref{fig2}d,e, confirm the validity of the above expectations for the 
BHH model (\ref{QMBHH3}) in the chaotic regime.

\begin{figure}[t]
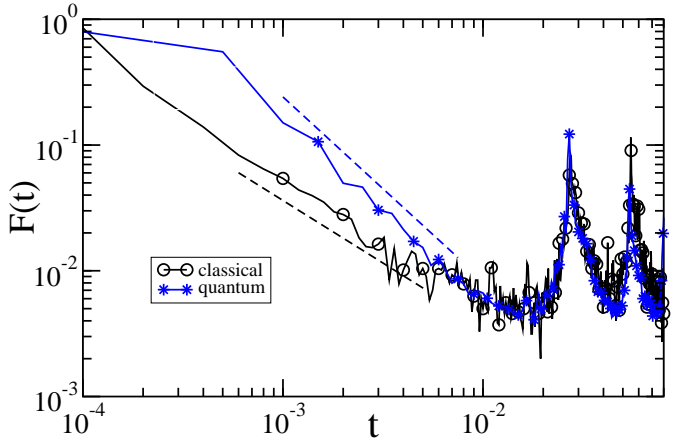

\onefigure[width=\columnwidth,keepaspectratio,clip]{fig3}
\caption{(Color online) The quantum (blue line) fidelity decay $F(t)$ at ${\tilde E}_n^{(0)}=0.39$, 
plotted together with the corresponding classical (black line) fidelity decay $F_{\rm cl}(t)$, calculated 
from eq.~(\ref{clfidel}). The perturbation strength is $\delta k =7.5>\delta k_{\rm prt}$. The dashed
lines are the best least square fits and are drawn to guide the eye. The upper one corresponds to 
$t^{-1.46}$ while the lower one corresponds to $t^{-0.99}$.
}
\label{fig3}
\end{figure}

Less clear is the situation for (predominantly) integrable systems for perturbation strengths 
$\delta k > \delta k _{\rm prt}$: depending on the initial state, the fidelity decays either 
faster than Gaussian (when the perturbation changes the frequencies of the phase-space tori) \cite{PZ02}, 
or in a power law fashion (when the primary effect of the perturbation is to change the shape 
of the phase-space tori) \cite{JAB03}. Indeed, our numerical results for the BHH in the (predominantly) 
integrable regime (high energies) confirm the above expectation (see figs.~\ref{fig2}f, \ref{fig3} 
and related discussion below). 

%----------------------------------------------------------------------------------------------------
%----------------------------------------------------------------------------------------------------

%%%%%%%%%%%%%%%%%%%    Fidelity echoes  %%%%%%%%%%%%%%%%%%%%%%%%%%%%%%%%%%

Next, we focus on the appearance of fidelity echoes. In order to gain some insight we turn first 
to the analysis of the perturbation matrix ${\bf B}$ which generates the dynamics. In 
figs.~\ref{fig2}a-c we show the band-profile $\langle |{\bf B}_{nm}|^2\rangle_E$ (dashed line) for 
various energy regimes $E$. One observes that the band-profile is not flat, but exhibits pronounced 
structures within the bandwidth $\Delta_b$. A striking feature is the existence of side-bands 
whose position $\omega_{\rm echo}$ increases as we increase the energy $E$. As a result $C(\tau)$ 
(see eq.~(\ref{LRT})) oscillates, leading to strong fidelity echoes at multiples of a characteristic 
time $t_{\rm echo}=2\pi/\omega_{\rm echo}$ (see figs.~2d-f). These echoes are different from the 
standard mesoscopic echoes at the Heisenberg time $t_{\rm H}=1/\Delta$ of quantum systems with 
chaotic classical dynamics~\cite{SS05}.They are instead associated with "non-universal" structures 
that dominate the band-profile of the perturbation matrix ${\bf B}$ and are the fingerprints of 
the lattice confinement. Quantum mechanically, these are reflected in selection rules that 
determine to which states $\ket{m(k_0\pm\delta k)}$ of $H_{1,2}$ an initial state $\ket{\Psi_0}$ is 
coupled. 

We consider the case where the initial preparation $\ket{\Psi_0}=\ket{n(k_0)}$ is an eigenstate 
of $H_0$ corresponding to an eigenvalue $E_n^{(0)}$. Here, the information about the coupling of 
the initial states to the perturbed states $\ket{m(k_0\pm\delta k)}$, is encoded in the structure 
of LDoS $\rho_L(E)=\sum_m |\langle m|n \rangle|^2 \delta[E-(E_m-E_n)]$. For $\delta k \leq \delta 
k_{\rm prt}$ we have that $|\langle m|n\rangle|^2 \approx \delta_{n,m}+{\delta k^2 \ \langle|{\bf 
B}_{nm}|^2\rangle}/[{\Gamma^2+(E_n{-} E_m)^2}]$, with a width $\Gamma \approx (\sigma \delta k)^2/
\Delta\leq \Delta_b$ indicating the energy regime where most of the probability is contained 
\cite{HKG06}. For short times $t\leq \Gamma^{-1}$, this core will not have time to dephase/decay. 
Therefore, the fidelity will show large echo recoveries at the time the side-bands have acquired 
a phase of $2\pi$, if $t_{\rm echo}\ll \Gamma^{-1}$. The numerical data for $\delta k\leq \delta 
k_{\rm prt}$ reported in figs.~\ref{fig2}d-f (see dashed lines) confirm this prediction. We see 
that the echo efficiency (\textit{i.e.} the recovery level of $F(t)$) increases as $t_{\rm echo}$ becomes 
smaller~\cite{note2}. Arguing along the same lines as above, we are able to explain the reduction
of fidelity echoes in the non-perturbative regime, $\delta k>\delta k_{\rm prt}$, observed in
figs.~\ref{fig2}d-f (see solid lines). Here, the LDoS covers the whole bandwidth, spoiling (with 
respect to the perturbative cases) the echo efficiency. 

%----------------------------------------------------------------------------------------------------
%\begin{figure}[t]
%\onefigure[width=\columnwidth,keepaspectratio,clip]{fig3}
%\caption{(Color online) The quantum (blue line) fidelity decay $F(t)$ at ${\tilde E}_n^{(0)}=0.39$, 
%plotted together with the corresponding classical (black line) fidelity decay $F_{\rm cl}(t)$, calculated 
%from eq.~(\ref{clfidel}). The perturbation strength is $\delta k =7.5>\delta k_{\rm prt}$. The dashed
%lines are the best least square fits and are drawn to guide the eye. The upper one corresponds to 
%$t^{-1.46}$ while the lower one corresponds to $t^{-0.99}$.
%}
%\label{fig3}
%\end{figure}
%----------------------------------------------------------------------------------------------------

Similar echoes were experimentally observed in~\cite{AGKD04} for atoms in optical billiards. However,
in contrast to our case (see black lines in figs.~\ref{fig2}d-f), Ref.~\cite{AGKD04} was reporting 
a total loss of echoes in the semiclassical regime $\delta k \ge \delta k_{\rm prt}$. To understand 
the origin of the echoes observed for the BHH model (\ref{QMBHH3}) for $\delta k \ge \delta k_{\rm 
prt}$, we will employ semiclassical considerations. In fig.~\ref{fig3}, we compare the quantum 
calculations for $\delta k>\delta k_{\rm prt}$ with the classical fidelity 
\begin{equation}
\label{clfidel}
F_{\rm cl}(t)= \int_{\Omega} d\textbf{x} \; 
\rho_{-\delta k}(\textbf{x},t) \rho_{+\delta k}(\textbf{x},t)
\end{equation}
where $\rho_{\pm \delta k}$ is the classical density function evolved under ${\cal H}(k_0\pm\delta 
k)$ and the integration is performed over the whole phase space $\Omega$~\cite{note4}. Although the quantum 
and classical calculations - as far as the echoes are concerned - agree quite well, a discrepancy 
on the fidelity decay between $F(t)$ and $F_{\rm cl}(t)$ for small times is also evident. In 
contrast to the chaotic regimes (lower energies), where for $\delta k>\delta k_{\rm prt}$ the 
fidelity decay was exponential with a rate given by the classical Lyapunov exponent (see for 
example figs.~2d,e), here $F_{\rm cl}(t) \sim t^{-\alpha}$ while the quantum fidelity decays
as $F(t)\sim t^{-3\alpha/2}$. The power law decay is a signature of classically (predominantly) 
integrable dynamics while the anomalous (faster) quantum decay is a pure quantum phenomenon, 
as was pointed out in \cite{JAB03}. According to the prediction of \cite{JSB01}, the classical 
power law exponent had to be $\alpha=d$, where $d=3$ is the dimensionality of the system (in 
fact, in the case of the BHH model (\ref{BHHCL}) we have a six-dimensional phase space with 
two constants of motion, namely the total number of particles $N$ and the energy $E$ and thus
$d_{\rm eff}=2$). However,
in the high energy regime, the dynamics is dictated by self-trapping phenomena, leading to 
localization of particles in one site. Therefore the effective dimensionality of the system
described by the Hamiltonian (\ref{BHHCL}) is $d_{\rm eff}=1$. Indeed, the best linear fit 
to the numerical data yields $\alpha = 1$, thus confirming the above argument. 
%----------------------------------------------------------------------------------------------------
%----------------------------------------------------------------------------------------------------

%----------------------------------------------------------------------------------------------------
\begin{figure}[t]
\onefigure[width=\columnwidth,keepaspectratio,clip]{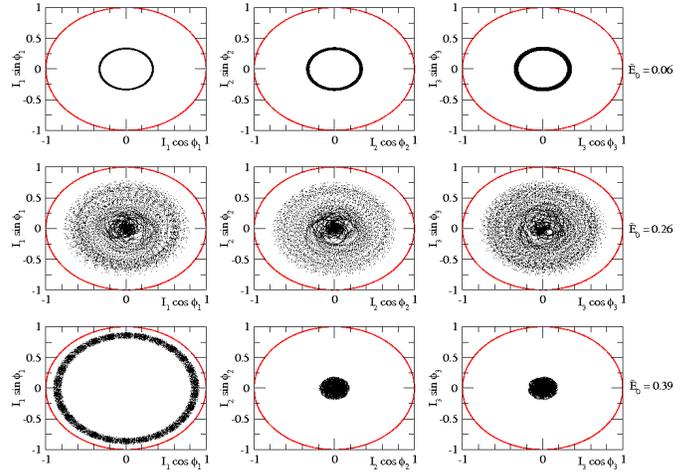}
\caption{(Color online) Time-evolution of $N=230$ particles (bosons) associated with various 
energy configurations: (top) ${\tilde E}=0.06$ close to the ground state (equipartition) of the 
system; (center) ${\tilde E}= 0.26$ corresponding to a state in the middle of the energy spectrum 
and (bottom) ${\tilde E}=0.39$ associated with a high energy state. The red line indicates the accessible 
phase space if all particles were on the corresponding site. Note that as the energy increases, 
the likelihood of "trapping" particles, \textit{i.e.}
particles that are initially localized in one site (in the specific case $i=1$) and stay 
there for long times, increases.
}
\label{fig4}
\end{figure}
%----------------------------------------------------------------------------------------------------

Let us now return to the analysis of the classical echoes. The classical trajectories contributing 
to the ensemble average of the echo, are those that -- after evolving forward in time with 
${\cal H}_1$ and then backwards in time with ${\cal H}_2$ -- return to the vicinity of their 
initial position. Since ${\cal H}_1$ and ${\cal H}_2$ are different in the coupling between 
nearby wells, those classical trajectories that do not jump between wells will not feel the 
difference in the coupling terms. Therefore they retrace their forward propagation backwards 
in time, causing the action integral to vanish. These trapped trajectories give a perfect 
contribution to the echo signal. Their existence is due to both the discreteness and the 
nonlinearity of the underlying equations of motion. For high enough energies the 
configuration of bosons, localized in one lattice site persists for long times as can be seen 
from fig.~\ref{fig4}. In such cases the bosons are said to be "self-trapped". As a result, 
the echo efficiency is increased in agreement with the numerical simulations presented in 
figs.~\ref{fig2}d-f. 

%----------------------------------------------------------------------------------------------------
\begin{figure}[t]
\onefigure[width=\columnwidth,keepaspectratio,clip]{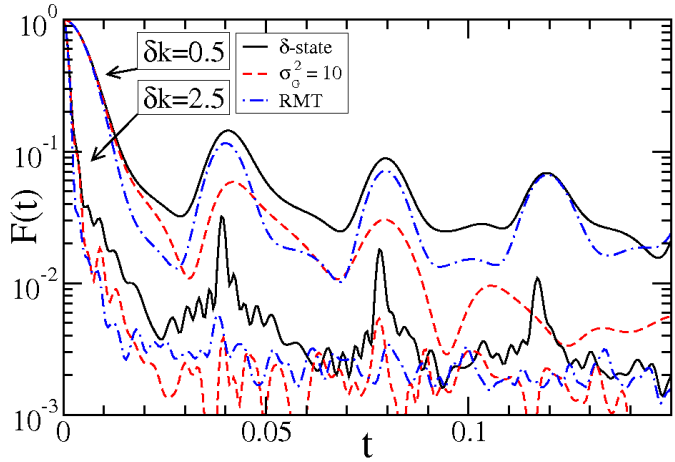}
\caption{(Color online) The fidelity $F(t)$ at ${\tilde E}_n^{(0)}=0.26$ for two perturbation 
strengths (upper curves $\delta k=0.5<\delta k_{\rm prt}$, lower curves $\delta k =2.5>\delta 
k_{\rm prt}$) and various initial conditions: black solid lines correspond to $\delta$ states, 
dashed lines to Gaussian states with variance $\sigma^2=10$, and dash-dotted lines to 
$\delta$-states evolved under the IRMT model. We have found that within the perturbative regime, 
the IRMT calculations coincide with the Linear Response Theory expression from eq.~(\ref{LRT}). }
\label{fig5}
\end{figure}
%----------------------------------------------------------------------------------------------------
We further confirm the importance of the structural average band-profile in the appearance of 
echoes, by evaluating the fidelity decay $F(t)$ using an improved RMT (IRMT) model~\cite{CK01}. 
While traditional RMT models involve band-profiles with matrix elements given by a Gaussian
distribution with a constant variance $\sigma^2$ within the bandwidth~\cite{SS05,GPSZ06}, the 
IRMT model incorporates the band-structure associated with the dynamical system, through the 
semiclassical relation (\ref{pspect}). In fig.~\ref{fig5} we present the results of these 
simulations. An excellent agreement with the fidelity calculations of the BHH (\ref{QMBHH3}) 
for the perturbative regime $\delta k < \delta k_{\rm prt}$ is evident. At the same time the 
IRMT modeling cannot describe the quantum results in the non-perturbative regime. Here, the 
echoes are related to subtler correlations of dynamical nature (self-trapping), which goes 
beyond the autocorrelation function $C(\tau)$ that determines the band-profile. 

Finally, in fig.~\ref{fig5} we report our numerical calculations for the case where $\ket{\Psi_0}$ 
is a superposition of eigenstates of $H_0$ with a Gaussian weight around some energy $E_0$. We 
found that as the width $\sigma_{G}^2$ of the Gaussian preparation increases, such that the 
participating eigenstates of $H_0$ span an energy window which is wider than $\delta E$ over 
which the band-profile of ${\bf B}$ is constant (see for example Figs.~\ref{fig2}a-c), the 
echo-efficiency is suppressed. This is due to destructive interferences resulting from states 
located at various parts of the spectrum where the band-structure (thus the $\omega_{\rm echo}$) 
is different. The destruction of echo efficiency becomes more pronounced in the non-perturbative
regime where no sign of echo is observed (see fig.~\ref{fig5}). 

Although the above analysis is focused on the $f=3$ case, the reported results are expected 
to be applicable for larger lattices $f>3$ as well. Furthermore, due to the connection between 
fidelity decay and loss of coherence \cite{CDPZ03}, our findings can be used to engineer coherence 
echoes of a central system coupled to a quantum bath consisting of cold atoms in an optical lattice
~\cite{AGKD04,RCGMF06}. 

We acknowledge T. Geisel for his continuous interest and support of this project. Useful
discussions with D. Cohen, N. Davidson, T. Gorin, F. Izrailev, Ph. Jacquod and D. Wright 
are also acknowledged.


\begin{thebibliography}{}
\bibitem{note3}
In fact, the perturbation-independent decay of $F(t)$ appears only for some 
special sets of initial states, like narrow wave packets \cite{CLMPV02,JSB01}. This has to be 
contrasted with the perturbative regime $\delta k < \delta k_{\rm prt}$, where the $F(t)$-decay 
is qualitatively the same for any initial preparation.
\bibitem{note2} 
We have checked that in the cases presented in Figs. \ref{fig2}d),e),f) the width $\Gamma$ 
remains approximately the same for corresponding $\delta k$-values.
\bibitem{note4} 
In the actual calculation the integration is done over an energy shell of width $dE$. 
We have checked that the numerical procedure is stable with respect to $dE$.
\end{thebibliography}
\end{document}